\documentclass[preprint2]{aastex}
\usepackage{longtable}
\usepackage{rotating}
\begin{document}

\title{ New \textit{HST} WFC3/UVIS observations augment the
stellar-population complexity of $\omega$~Centauri\footnote{Based on
observations with the NASA/ESA \textit{Hubble Space Telescope},
obtained at the Space Telescope Science Institute, which is operated
by AURA, Inc., under NASA contract NAS 5-26555.
\newline
$^{**}$ Visiting PhD Student at STScI
    under the \textit{``2008 graduate research assistantship''}
    program.}}

\author{
Bellini, A.\altaffilmark{2,3,**},
Bedin, L. R.\altaffilmark{3},
Piotto, G.\altaffilmark{2},
Milone, A. P.\altaffilmark{2},
Marino, A. F.\altaffilmark{2,4},
Villanova, S.\altaffilmark{5}
}
\altaffiltext{2}{Dipartimento di Astronomia, Universit\`a di Padova,
vicolo dell'Osservatorio 3, I-35122 Padova, Italy, EU}
\altaffiltext{3}{Space Telescope Science Institute, 3700 San Martin
Drive, Baltimore, MD 21218, USA}
\altaffiltext{4}{P. Universidad Catolica de Chile, Departamento de
Astronomia y Astrofisica, Casilla 306, Santiago 22, Chile}
\altaffiltext{5}{Departamento de Astronomia, Universidad de
Concepcion, Casilla 160-C, Concepcion, Chile}

\email{bellini@stsci.edu}

\begin{abstract}
{We used archival multi-band {\it Hubble Space Telescope} observations
obtained with the {\it Wide Field Camera 3} in the {\it UV-optical}
channel to present new important observational findings on the
color-magnitude diagram (CMD) of the Galactic globular cluster
$\omega$~Centauri. The ultraviolet WFC3 data have been coupled with
available WFC/ACS optical-band data. The new CMDs, obtained from the
combination of colors coming from eight different bands, disclose an
even more complex stellar population than previously identified. This
paper discusses the detailed morphology of the CMDs.}
\end{abstract}


\section{Introduction}
\label{introduction}

No doubt, $\omega$ Centauri is the most-studied and the most-enigmatic
among the Milky Way satellites.  For long times it has been considered
a Globular Cluster, but a number of peculiarities, like the mass, the
chemical composition, the stellar content and the kinematics, suggest
that it might be the remnant of a larger stellar system (Bekki \&
Freeman 2003, Lee et al.\ 2009, and references therein).

Great interest and great efforts have been dedicated to this object,
since the discovery that its stars span a wide range of metallicities,
including iron-peak elements (Cannon \& Stobie 1973, Freeman \&
Rodgers 1975, Johnson et al.\ 2009 and references therein).

With the advent of wide-field imagers and thanks to the increasingly
high-photometric precision in the densest cluster regions, new
discoveries revived the interest in $\omega$~Cen, and surely
complicated the already inexplicable enigma represented by its
composite stellar population.  Lee et al.\ (1999) and Pancino et al.\
(2000) announced that its red giant branch (RGB) resolves into several
distinct stellar sequences.  Anderson (1997) found that, over a range
of about two magnitudes, the main sequence (MS) splits into a blue
(bMS) and a red sequence (rMS). The result has been confirmed by Bedin
et al.\ (2004), who discovered a third, less populated MS (MS-a) on
the red side of the rMS (see also Villanova et al.\ 2007, hereafter
V07).  A totally unexpected discovery came from the spectroscopic
analysis by Piotto et al.\ (2005), who revealed that the bMS is more
metal-rich than the rMS.  Only greatly-enhanced helium can explain the
color and metallicity difference between the two MSs.  Bellini et al.\
(2009) (see also Sollima et al.\ 2007), found that bMS stars are more
centrally concentrated than rMS ones, with a bMS over rMS ratio
ranging from $\sim$1.0 ($r$$\lesssim$$2\farcm5$) to $\sim$0.40
($r$$\gtrsim$8$\arcmin$).

\begin{figure*}[!t]
\centering
\includegraphics[width=16.5cm]{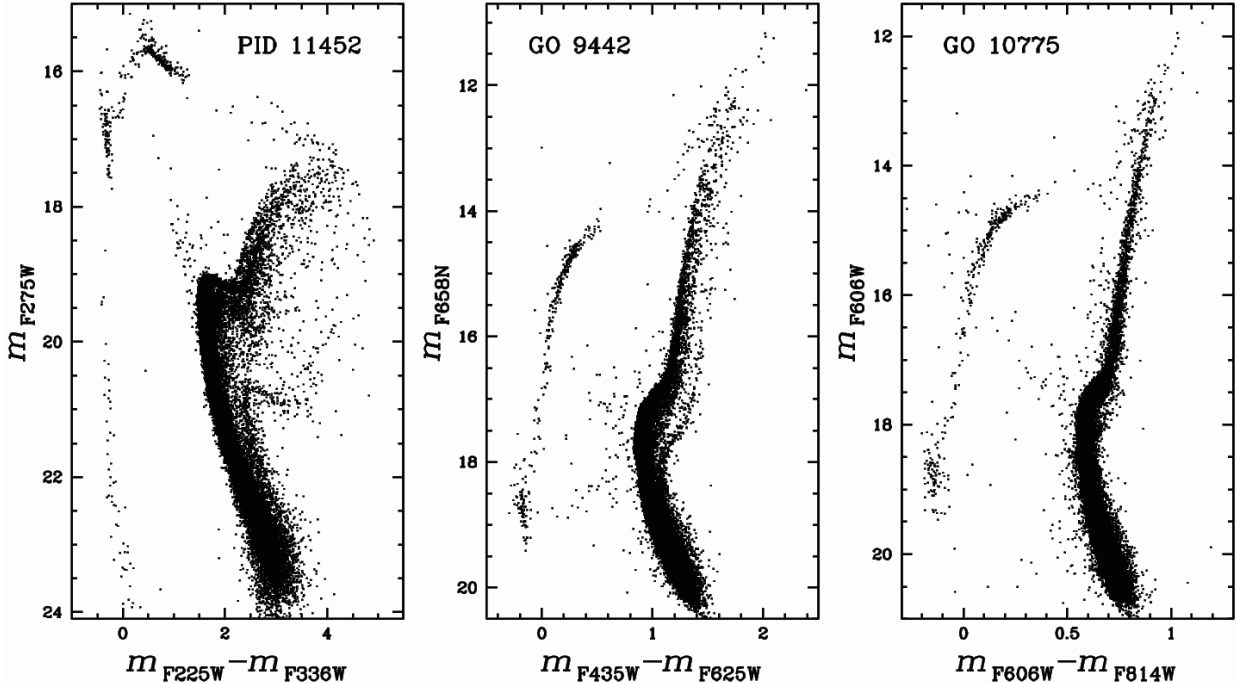}
\caption{A collection of CMDs from our three data sets: the WFC3/UVIS
  CMD in the UV filters $m_{\rm F275W}$ vs.\ $m_{\rm F225W}-m_{\rm
    F336W}$, from PID-11452 (\textit{left}); the ACS/WFC $m_{\rm
    F658N}$ vs.\ $m_{\rm F435W}-m_{\rm F625W}$ CMD, from GO-9442
  (\textit{center}); and the ACS/WFC $m_{\rm F606W}$ vs.\ $m_{\rm
    F606W}-m_{\rm F814W}$ CMD, from GO-10775 (\textit{right}). We
  plotted only the best $\sim 32\,000$ stars in common among the three
  data sets (see text for details).}
\label{fig1}
\end{figure*}

\begin{figure*}[!t]
\centering
\includegraphics[width=16.5cm]{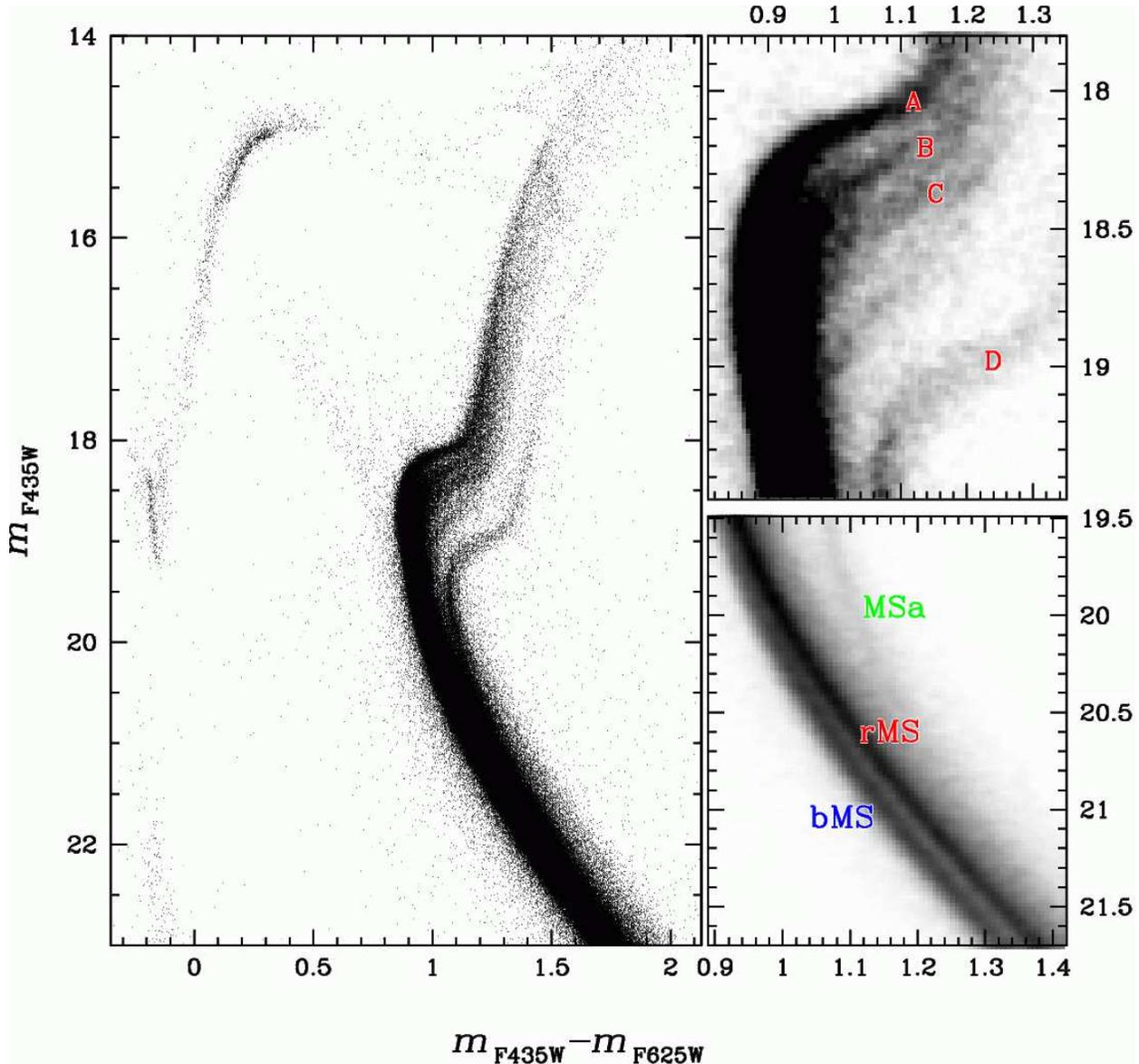}
\caption{Definition of the main CMD branches used in the present
paper.}
\label{fig2}
\end{figure*}

\begin{figure*}[!t]
\centering \includegraphics[width=16.5cm]{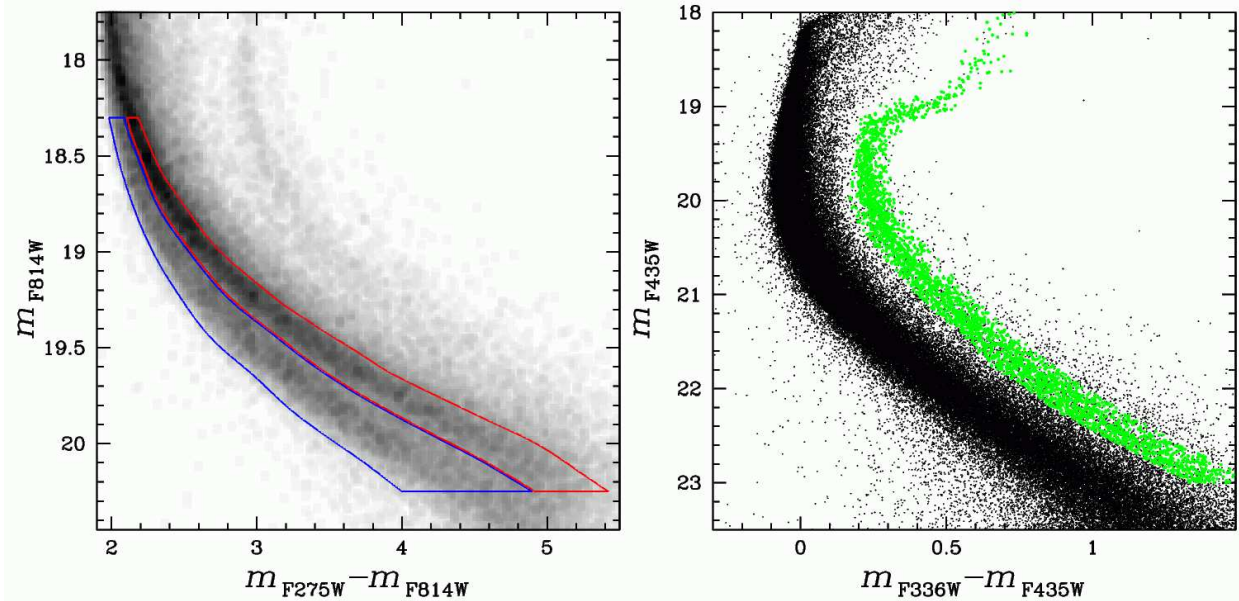}
\caption{\textit{(Left)}: Selection of bMS (blue boundary) and rMS (red
    boundary) stars. \textit{(Right)}: Selection of MS-a (green
  points) stars.}
\label{fig3}
\end{figure*}

\begin{figure*}[!t]
\centering \includegraphics[width=16.5cm]{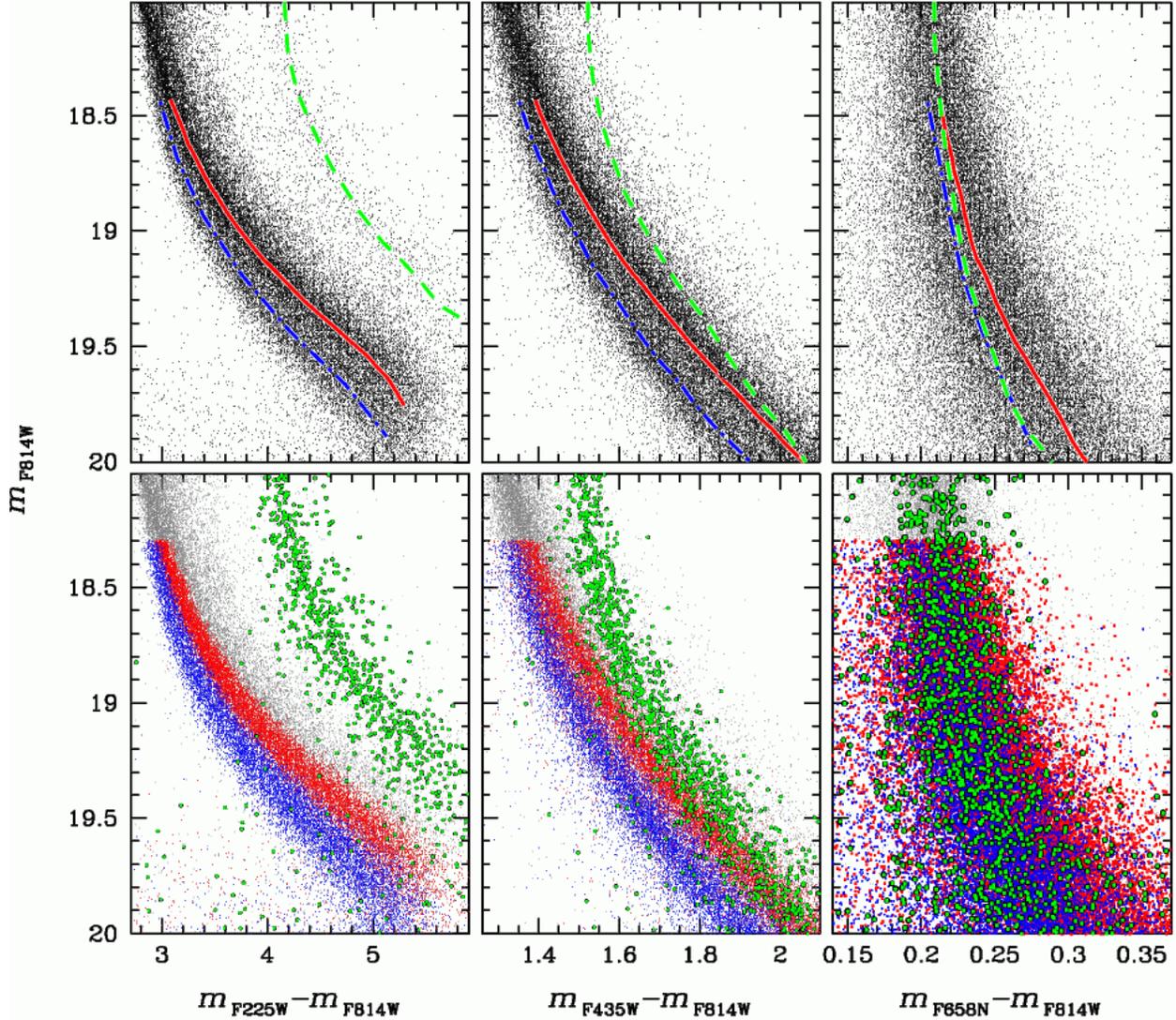}
\caption{Example of the definition of the MS RLs.}
\label{fig4}
\end{figure*}

\begin{figure*}[!t]
\centering \includegraphics[width=16.5cm]{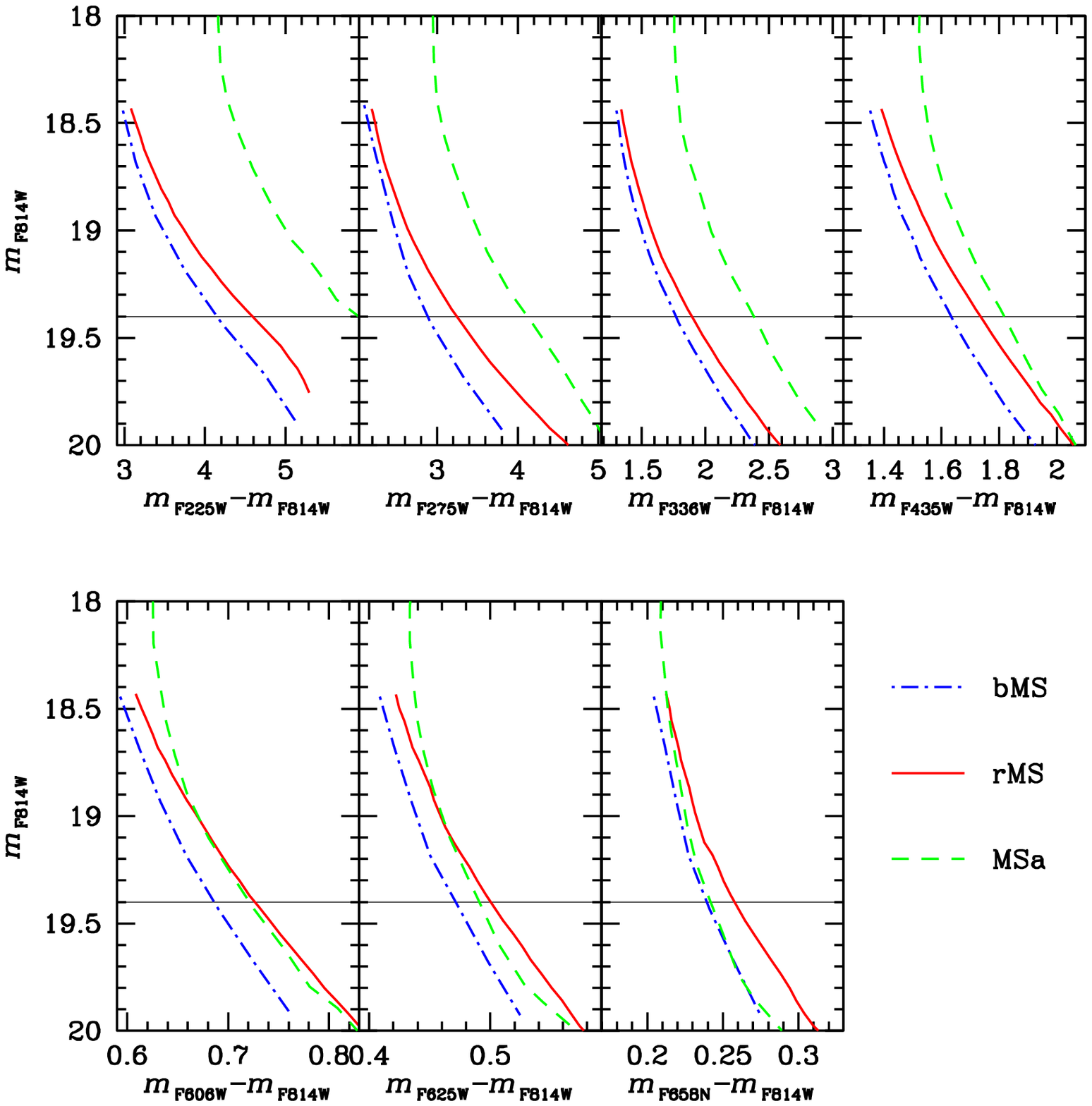}
\caption{The bMS and the MS-a run almost parallel in all CMDs. In some
CMDs the MS-a seems to cross the rMS.}
\label{fig5}
\end{figure*}

Perhaps, the most complex region of the color-magnitude diagram (CMD)
is the sub-giant branch (SGB).  Photometric studies showed that the
SGB of $\omega$~Cen is split into 4, possibly 5, distinct stellar
populations (Lee et al.\ 2005, Sollima et al.\
2005, V07).

In this work we present high-accuracy photometry obtained with both
the {\it Wide Field Camera 3} (WFC3) in the {\it UV-optical} channel
(UVIS), and the {\it Wide Field Channel} of the {\it Advanced
Camera for Survey} (WFC/ACS) of the {\it Hubble Space Telescope}
(\textit{HST}).  Not surprisingly, we obtained astonishingly complex
CMDs unveiling a number of new details which make the $\omega$~Cen stellar
population more complex than ever shown, making the the understanding
of the star formation history in this cluster a real challenge.

The purpose of this work is just to present to the astronomical
community these new CMDs, discuss their detailed morphology -- with
the hope of adding important information and a deeper insight into
$\omega$~Cen -- and help to add up all the pieces of what still
remains a broken puzzle.  The multi-band astro-photometric catalog
presented in this work will be publicly available to the astronomical
community, for further analysis.

\section{Observations, Measurements, and Selections}

For calibration purposes, $\omega$~Cen has been observed many times
with {\it HST}, using a huge variety of filters.  Recent observations
were collected with the newly installed WFC3.  In September 9$^{\rm
  th}$ 2009, a set of well-dithered exposures through the broad-band
ultraviolet (UV) filters F225W, F275W, and F336W were released to the
community.  The data were collected in July 15, 2009, for general
calibration purposes (PID-11452). The portion of the data that we used
in this paper consists of 9 exposures, each of 350 s, for each filter.
The archive images were standard pipe-line pre-reduced {\sf FLT}, and
we measured star positions and fluxes with a software mostly based on
{\sf img2xym\_WFI} (Anderson et al.\ 2006).  Details on
this software will be presented in a stand-alone paper.  Star
positions and fluxes have been corrected for geometric distortion and
pixel-area using the geometric distortion solution provided by Bellini
\& Bedin (2009), and calibrated as in Bedin et al.\ (2005).

We complement these UV data with the optical photometry obtained with
the ACS/WFC in the filters F435W, F625W, F658N, F606W, and F814W.
Details on these data sets and their photometry can be found in V07,
and Anderson et al.\ (2008).

Since we are focused on high-accuracy photometry, this work only
concerns relatively isolated stars with small photometric and
astrometric errors, and high PSF-fit quality. A detailed description
of the selection procedures adopted in this paper is given in Milone
et al.\ (2009).  Finally, we corrected our photometry for
both reddening variations in the field of view (FoV) and
spatial-dependent photometric errors, introduced by small variations
of the PSF shape, which are not accounted for in our PSF models.  With
this aim, we used a method similar to that used by Sarajedini et al.\
(2007) and Milone et al.\ (2008). Briefly,
we determined the average MS ridge line (RL) for each CMD and then we
analyzed the color residuals as a function of the position within the
FoV. We corrected the effect of spatial photometric variations
suffered by each star by computing the average color residuals from
the MS RL of its 50 well-measured neighbors, and by correcting the
star color by this amount.

\section{Color-Magnitude Diagrams}

Figure\ \ref{fig1} shows a collection of CMDs from multi-band
WFC3/UVIS and ACS/WFC photometry.  All CMDs encompass all the
evolutionary sequences, from faint MS stars, down to a well developed
white dwarf (WD) cooling sequence (CS, see Fig.~\ref{fig1}).  A close
view of these CMDs is sufficient to realize that each of them is a
mine of information on the stellar content of $\omega$~Cen.  A
model-based interpretation of these CMDs is severely complicated by
the heterogeneity of the composition of each sequence and by possible
age differences, and requires a very accurate analysis, which is
beyond the purpose of the present paper.

Many of the features that we observe in these CMDs are well known, and
widely studied. For completeness of information, and in order to make
the following discussion clearer to the reader, we show in
Fig.~\ref{fig2} (left panel) the CMD resulting from the $10 \times 10$
arcmin$^{2}$ mosaic of ACS images centered on the cluster center, that
was already analyzed in several papers (Bedin et al.\ 2004, V07,
Cassisi et al.\ 2009, Bellini et al.\ 2009, and D'Antona, Caloi, \&
Ventura 2010).  The high accuracy of the ACS photometry already
revealed a large number of evolutionary sequences in the CMD. We used
Hess diagrams on the right panels of Fig.\ \ref{fig2} to highlight the
four main SGBs and the triple MS, following the notation of V07.

In the following, we will focus our attention on a number of details
in the CMD that can be revealed for the first time by the
high-accuracy multi-band photometry presented in this paper.

\subsection{The triple main sequence}

The new, multi-band data set provided by WFC3, combined with the ACS
data, open a new observational window also on the complex main sequence
of $\omega$~Cen.

The wide color base-line of the $m_{\rm F814W}$ vs. $m_{\rm
F275W}-m_{\rm F814W}$ CMD plotted on the left panel of Fig.~\ref{fig3}
allows us to isolate the two groups of bMS and rMS stars indicated by
blue and red color-coded regions.  Similarly, we can select a sample
of MS-a stars from the $m_{\rm F435W}$ vs. $m_{\rm F336W}-m_{\rm
F435W}$ CMD where the MS-a is most clearly separated from the
remaining MSs of $\omega$~Cen. Selected stars are highlighted in green
on the right panel CMD of Fig.~\ref{fig3}.

We have high-accuracy photometric measurements in eight bands, which
allow us to plot seven distinct CMDs involving the F814W band. For
each of them, we plotted $m_{\rm F814W}$ magnitudes as a function of
the $m_{X}-m_{\rm F814W}$ color, where $m_{X}=m_{\rm F225W}$, $m_{\rm
  F275W}$, $m_{\rm F336W}$, $m_{\rm F435W}$, $m_{\rm F606W}$, $m_{\rm
  F625W}$ and $m_{\rm F658N}$.

The bottom panels of Fig.~\ref{fig4} show the three most
representative of these CMDs (zoomed around the MS region). We
assigned to each star a blue, red or green color code according to
whether it belongs to the bMS, rMS, or MS-a sample, as defined in
Fig.~\ref{fig3}.  In the upper panels of Fig.~\ref{fig4} we
overimposed to the observed CMDs the MS RLs corresponding to the three
MSs, extracted from the CMD using the method described in Milone et
al.\ (2008).  Briefly, we divided the CMD in intervals of
0.2 magnitudes in the F814W band and calculated for each of them the
median color and magnitude for the bMS, rMS, and MS-a stars.  We
fitted these median points with a spline and obtained a first guess
for the MS RL.  Then, we calculated the difference between the color
of each star and the color of the MS RL corresponding to magnitude of
the star, and we took as $\sigma$ the $68.27^{\rm th}$ percentile of
the absolute value of the color difference. We rejected all stars with
color differences larger than $4 \sigma$, and we recalculated the
median points.

The RLs for the three MSs are shown in Fig.~\ref{fig5} for the seven
CMDs analyzed in this section.  We note a few interesting features:
(i) the RLs of the bMS and the MS-a are nearly parallel in all the
CMDs in the magnitude range $m_{\rm F814W} \sim 18.5-20.0$, while the
RL of the rMS have a different slope; (ii) when using
$m_{F606W}-m_{\rm F814W}$ and $m_{F625W}-m_{\rm F814W}$ colors the RL
of the MS-a seems to intercept (or merge with) the rMS going from the
brightest to the faintest stars.  MS-a stars become even bluer than
the rMS ones in the $m_{\rm F658N}-m_{\rm F814W}$ color.

This is the most intriguing CMD, in the context of the He content:\ He
abundance affects the color of MS stars.  The F658N filter maps
essentially the H$\alpha$ feature, with a very small influence by
other elements. It measures the strength of the H$\alpha$ which, for
MS stars cooler than 8000 K, is a function of the T$_{\rm eff}$, but
also of the hydrogen content, if it is allowed to vary.  MS-a stars
are more metal rich than rMS stars, being the progenitors of SGB-D and
of the RGB-a (Pancino et al.\ 2002, V07).  For this
reason, the fact that MS-a stars become even bluer than the rMS in the
$m_{\rm F658N}-m_{\rm F814W}$ color, overlapping with the bMS, might
imply that also MS-a is enriched in He (as suggested by Norris
2004). In fact, He enhancement tends to move the MS to bluer
colors.  The shape of the the MS-a, parallel to the bMS, might be also
an indication that its stars are He enriched. However, we also know
that the MS-a has higher iron content than the bMS. Higher metallicity
implies redder MS colors.  It is a combination of different metal
abundances, including CNO, and He content which results in the
observed behavior of the MS-a color.

\begin{figure}[!t]
\centering
\includegraphics[width=\columnwidth]{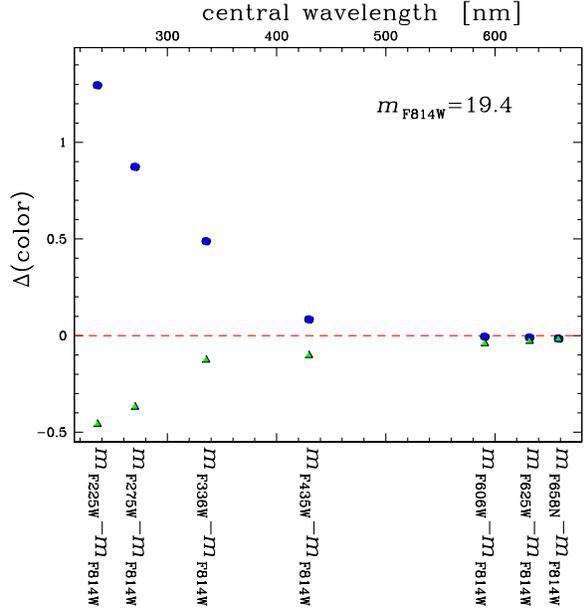}
\caption{Color distance from the rMS RL for bMS stars (blue points)
and MS-a stars (green triangles) at $m_{\rm F814W}=19.4$, plotted as a
function of the central wavelength of the $m_{\rm X}$ filter.  }
\label{fig6}
\end{figure}

\begin{table}[t!]
\caption{Color distances from the rMS RL for bMS stars ($\Delta_{\rm
COLOR}^{\rm bMS}$) and MS-a stars ($\Delta_{\rm COLOR}^{\rm MS-a}$) at
$m_{\rm F814W}=19.4$.}
\label{deltacolor}
\vskip 2mm
\centering 
\footnotesize{
\begin{tabular}{ccccc}
\hline\hline
COLOR&$\Delta_{\rm COLOR}^{\rm bMS}$&$\Delta_{\rm COLOR}^{\rm MS-a}$\\
\hline
$m_{\rm F225W}-m_{\rm F814W} $ & $  1.30\pm0.01$  & $ -0.46\pm.01$0 \\
$m_{\rm F275W}-m_{\rm F814W} $ & $  0.87\pm0.01$  & $ -0.37\pm.01$0 \\
$m_{\rm F336W}-m_{\rm F814W} $ & $  0.49\pm0.01$  & $ -0.13\pm.01$0 \\
$m_{\rm F435W}-m_{\rm F814W} $ & $  0.08\pm0.01$  & $ -0.10\pm.01$0 \\
$m_{\rm F606W}-m_{\rm F814W} $ & $ -0.01\pm0.01$  & $ -0.04\pm.01$0 \\
$m_{\rm F625W}-m_{\rm F814W} $ & $ -0.01\pm0.01$  & $ -0.03\pm.01$0 \\
$m_{\rm F658N}-m_{\rm F814W} $ & $ -0.02\pm0.01$  & $ -0.02\pm.01$0 \\
\hline
\hline
\end{tabular}}
\end{table}

In order to quantify the color differences among the three MSs as a
function of the color baseline, in Fig.~\ref{fig6} we plotted the central
wavelength $\lambda$ of the $m_{\rm X}$ filter versus the measured
color-difference $\Delta$(color) of both bMS stars (blue points) and
MS-a stars (green points), with respect to the rMS RL color, at
$m_{\rm F814W}=19.4$ (this magnitude level is also indicated with an
horizontal line in Fig.~\ref{fig5}).

The color distances plotted in Fig.~\ref{fig6} are listed in
Table~\ref{deltacolor}.

\subsection{The intrinsic broadening of the rMS}
\label{sec:intrinsic_broad_rMS}

\begin{figure*}[!t]
\centering
\includegraphics[width=16.5cm]{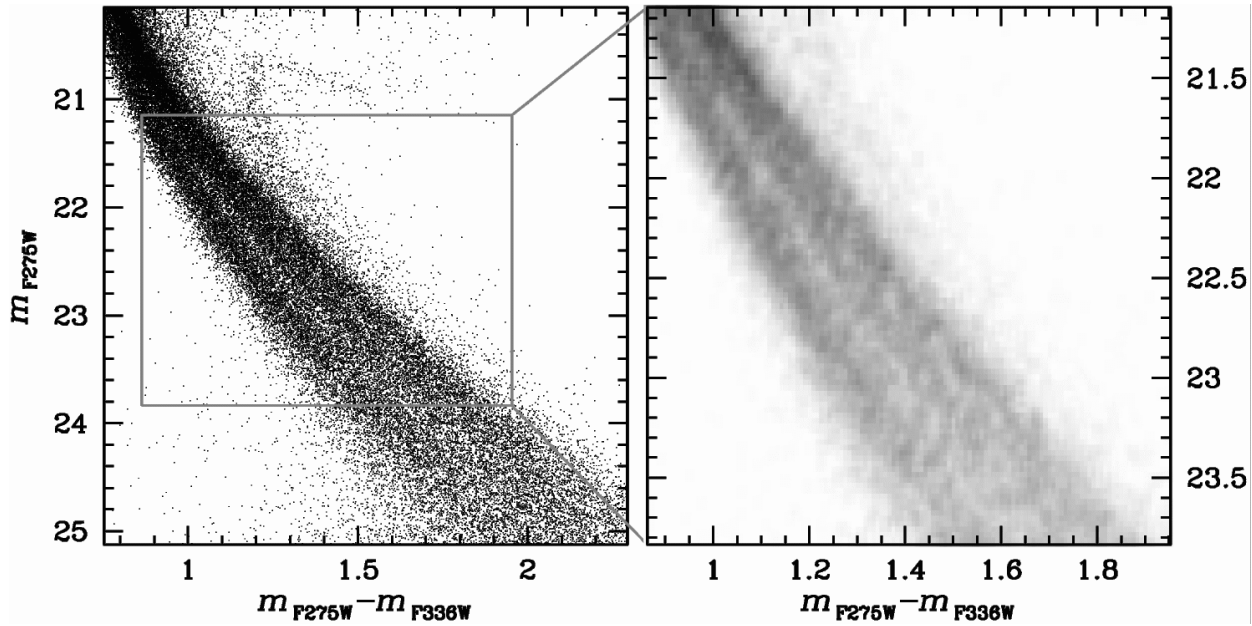}
\caption{The rMS is broadened (see text for details).}
\label{fig7}
\end{figure*}

A visual inspection of the $m_{\rm F275W}$ vs.  $m_{\rm F275W}-m_{\rm
  F336W}$ CMD of Fig.~\ref{fig7} suggests that the rMS is broadened.
In this section we will investigate the possible presence of this
intrinsic color spread among rMS stars, by using the same approach
followed in the recent studies on the MS broadening of 47 Tuc
(Anderson et al.\ 2009) and of NGC 6752 (Milone et
al.\ 2010).

\begin{figure*}[!t]
\centering
\includegraphics[width=16.0cm]{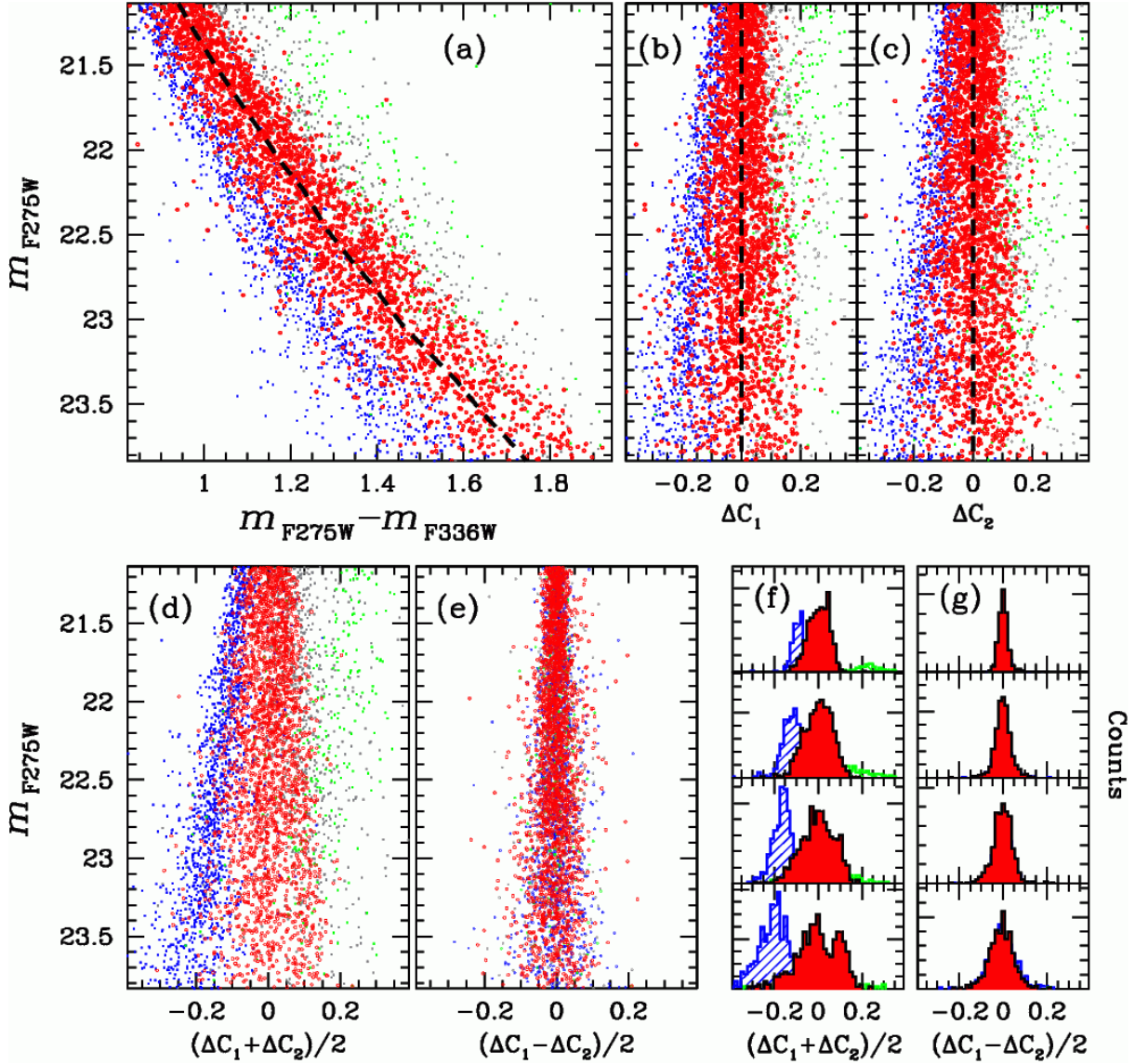}
\caption{Evidence of the intrinsic broadening of the rMS (see text for
  details).}
\label{fig8}
\end{figure*}

We started by dividing F275W and F336W images in two halves
(hereafter, samples 1 and 2), and considered only those stars measured
in both sub-samples.  From each of the two independent sub sets, we
plotted a CMD.  As an example we show the CMD from the first data set
in panel (a) of Fig.~\ref{fig8}.  The selected bMS and MS-a stars are
represented with blue and green colors, while rMS stars are plotted in
red (in all the plots, star colors are given according to their
classification, as defined in Fig.~5).  The dashed line is the RL of
the rMS, obtained as described in the previous section.  Then we
subtracted from the observed color (hereafter $C$) of each star the RL
color at the same magnitude, obtaining the quantity $\Delta C$.  The
straightened MSs for the first and the second data set are plotted in
panels (b) and (c), respectively.  In panel (d) we show the color
distribution of the straightened MS from the whole data set [indicated
  as $(\Delta C_{1}+\Delta C_{2})/2$].  In this case, the errors are
smaller by a factor $\sqrt{2}$ with respect to those of the two data
halves.

Panel (e) shows the distribution of the difference between the colors
in each half of the images [i.e., $(\Delta C_{1}-\Delta C_{2})/2$]
which is indicative of the color error.  The histogram distribution of
$(\Delta C_{1}+\Delta C_{2})/2$ and $(\Delta C_{1}-\Delta C_{2})/2$
are plotted in panels (f) and (g), respectively.  In
Table~\ref{tabdata} we give the estimated values of the intrinsic and
error dispersion of the rMS, for four equally-spaced magnitude
intervals, assuming a Gaussian distribution.  As suggested by a visual
inspection of panel (d), there is no doubt that the rMS is larger than
what expected from the color-error distribution shown in panel
(e)\footnote{Note that we are aware of the significantly different
  efficiencies of the two CCDs of WFC3/UVIS toward UV. For this
  reason, thanks to the large dither pattern of the observations, we
  were able to repeat the analysis creating two subsamples made up
  with only one, or the other, CCD.  We find (in the smaller region of
  the overlap) the same color distribution for the rMS.}. Even in the
worst case of the last considered magnitude bin, we have an intrinsic
dispersion of $0.100\pm0.004$, which is significantly larger (at the
level of more than 10$\sigma$) than the error dispersion
($0.051\pm0.002$).

\begin{table}[t!]
\centering
\caption{The two quantities representing estimates of the intrinsic dispersion
(second column), and of our measurements uncertainties (third column),
in four different magnitude intervals (indicated in the first column). }
\label{tabdata}
\vskip 2mm
\footnotesize{
\begin{tabular}{ccccc}
\hline
\hline
  $m_{\rm F275W}$ & $\sigma_{(\Delta C_{1}+\Delta C_{2})/2}$
                & $\sigma_{(\Delta C_{1}-\Delta C_{2})/2}$ \\
\hline
  21.13--21.81 &  $0.048\pm0.002$ & $0.021\pm0.001$ \\
  21.81--22.48 &  $0.067\pm0.002$ & $0.033\pm0.001$ \\
  22.48--23.16 &  $0.082\pm0.003$ & $0.040\pm0.002$ \\
  23.16--23.83 &  $0.100\pm0.004$ & $0.051\pm0.002$ \\
\hline
\end{tabular}}
\end{table}

\begin{figure*}[!t]
\centering
\includegraphics[width=16.5cm]{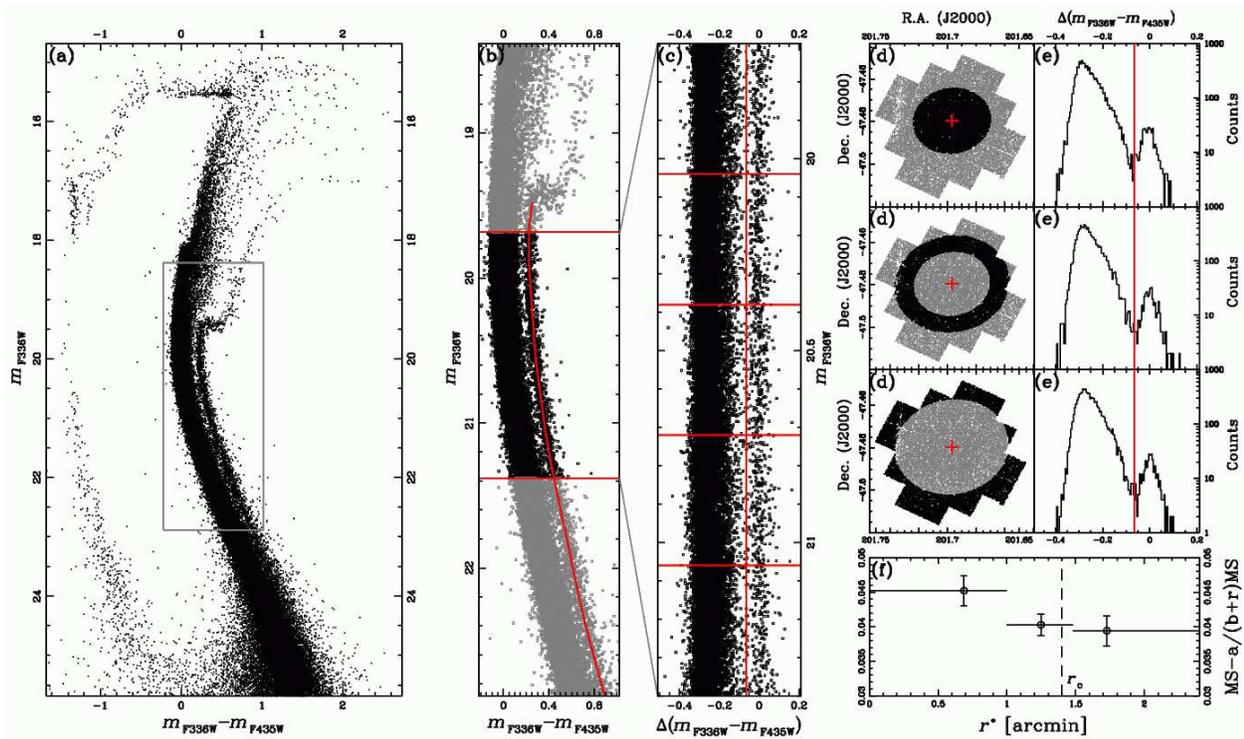}
\caption{Panel (a) shows the $m_{\rm F336W}$ vs.\ $m_{\rm
    F336W}-m_{\rm F435W}$ CMD.  The region highlighted with the grey
  rectangular is zoomed-in in panel (b). The red fit marks the MS-a
  fiducial line.  In panel (c) we show the rectified MSs in the
  magnitude interval 19.7$<$$m_{\rm F336W}$$<$21.4. The vertical red
  line separates MS-a members (on the right) from the rest of MS stars
  (on the left). We defined 3 radial bins [panels (d)], each one
  containing the same number of selected stars. For each radial bin we
  derived a color-distribution histogram [panels (e)]. The radial
  distribution of the MS-a/(b+r)MS star-count ratio is shown in panel
  (f).  The dashed line marks the core radius (Harris
  1996).}
\label{fig9}
\end{figure*}

Figure~\ref{fig7} and the bottom two panels of Col.\ (f) of
Fig.~\ref{fig9} might suggest a possible split. We cannot assess the
significance of this feature, but we think it is worth of further
investigation.  As shown by V07, the rMS evolves into the brightest
SGB-A sequence.  The fact that the rMS is broadened shall not come as
a surprise. In Piotto et al. (2005), stars in this sequence
were found to have a large spread in C, much larger than that of stars
on the bMS.  If light-element abundances are correlated, as it happens
in all the massive clusters, this implies also a spread in N and O. We
do not know the magnitude of this spread, but it surely must be
reflected in the photometry, especially in the blue-UV filters where
CH and CN bands are located.

\begin{figure*}[!t]
\centering
\includegraphics[width=16.0cm]{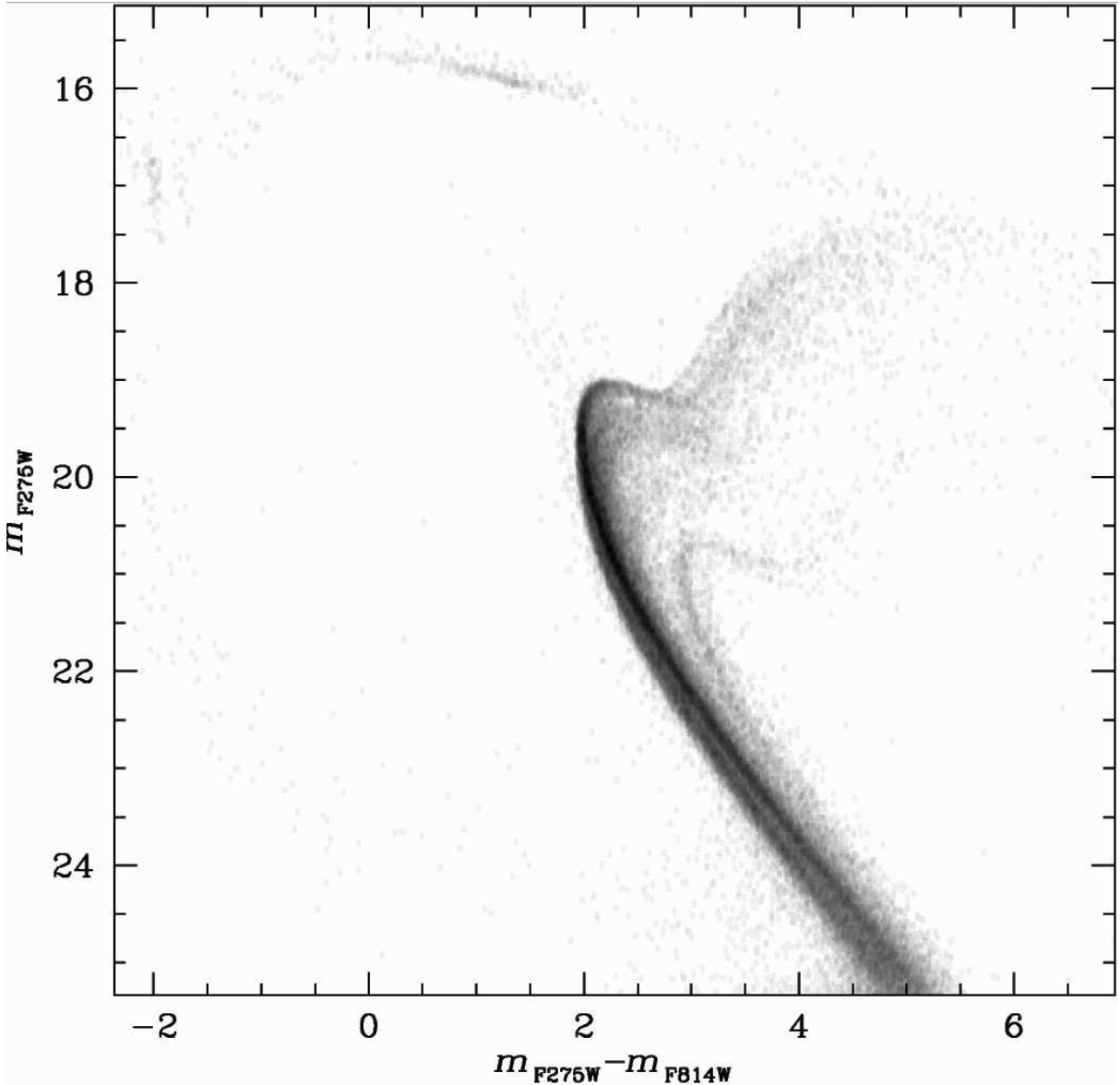}
\caption{Hess diagram of the $m_{\rm F275W}$ vs.\ $m_{\rm
    F275W}-m_{\rm F814W}$ CMD showing the complexity of $\omega$~Cen.}
\label{fig10}
\end{figure*}

\begin{figure*}[!t]
\centering
\includegraphics[width=15cm,height=16.0cm]{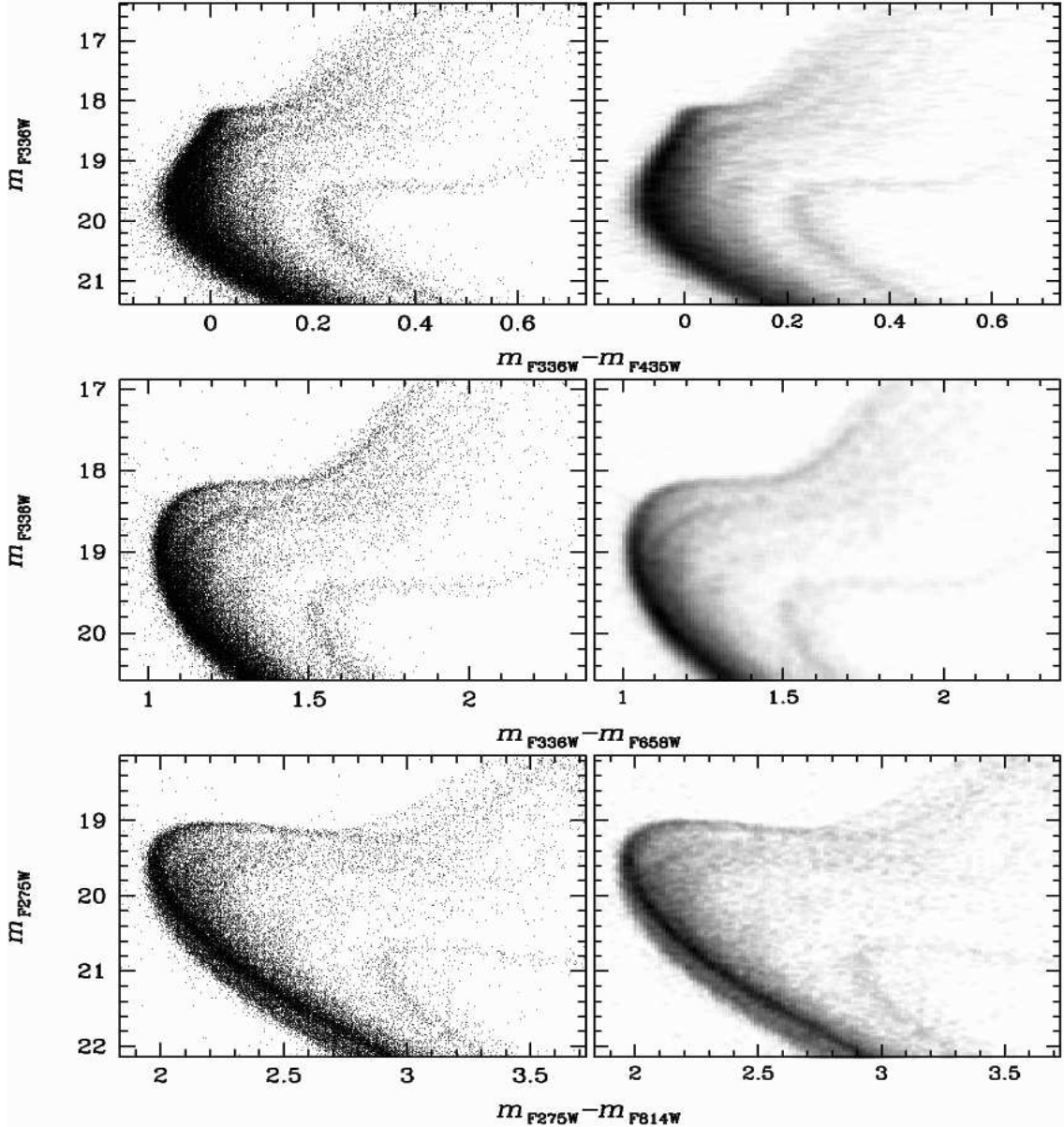}
\caption{CMDs and corresponding Hess diagrams in different bands,
focused around the SGBs region.}
\label{fig11}
\end{figure*}

\begin{figure*}[!t]
\centering
\includegraphics[width=14cm]{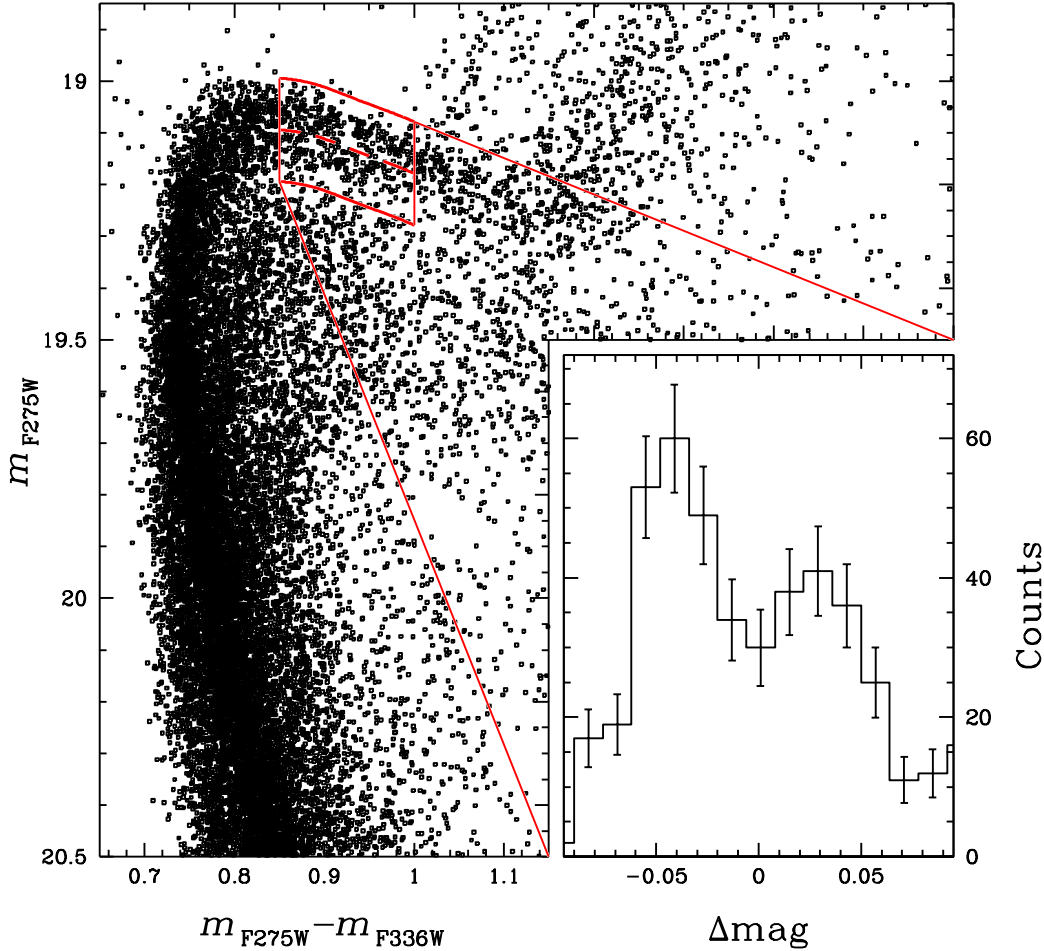}
\caption{Zoom-in of the SGB in the $m_{\rm F275W}$ vs.\
 $m_{\rm F275W}-m_{\rm F336W}$ CMD. On the lower right, the histogram
 of the magnitude difference between the stars inside the red box and
 the magnitude of the dashed line, at the same color of the stars (see
 text for more details).}
\label{fig12}
\end{figure*}

\begin{figure*}[!t]
\centering
\includegraphics[width=16.0cm]{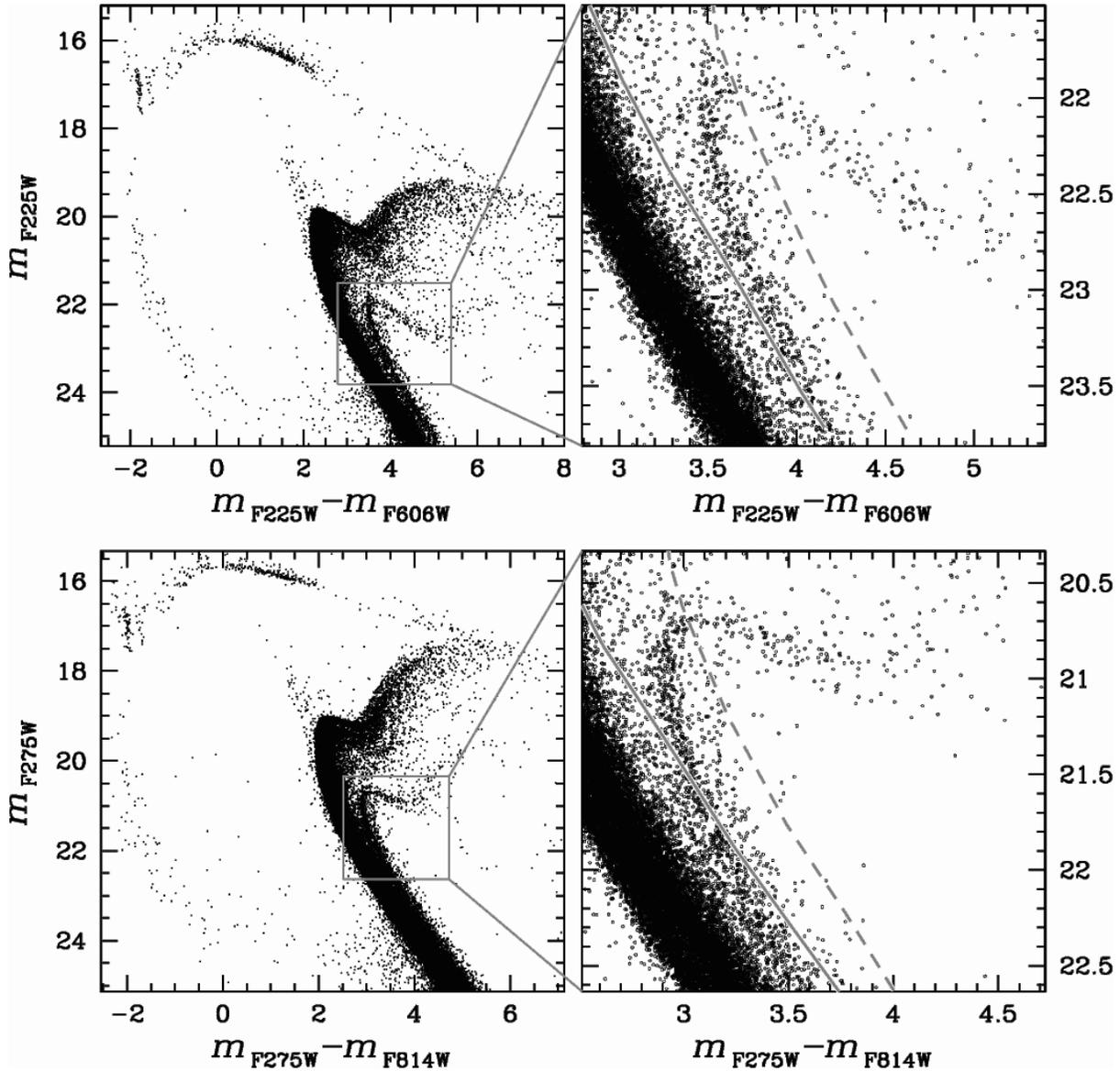}
\caption{Zoom-in of two CMDs around the SGB-D region showing hints of
 two sub-groups.}
\label{fig13}
\end{figure*}

\begin{figure*}[!t]
\centering
\includegraphics[width=16.0cm]{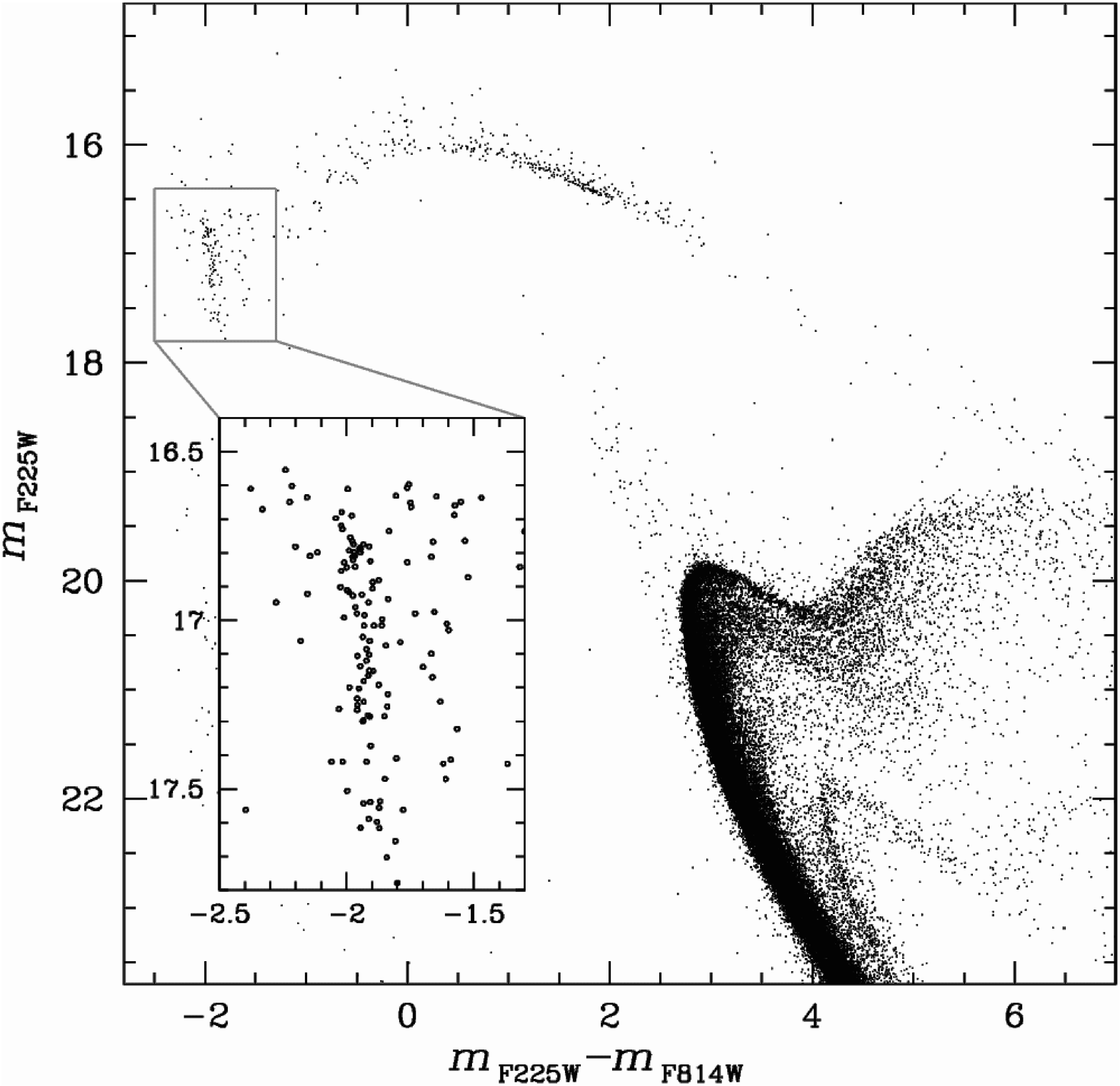}
\caption{The $m_{\rm F225W}$ vs.\ $m_{\rm F225W}-m_{\rm F814W}$ CMD
best highlights the complex morphology of the HB. The inset shows a
zoom-in of the blue hook section of the HB. Two distinct and almost
parallel features are visible.}
\label{fig14}
\end{figure*}

\subsection{The MS-a.}

The accurate CMDs presented in previous sections impose a more
detailed investigation also for the MS-a.

Among all color combinations, the $m_{\rm F336W}-m_{\rm F435W}$ color
is the one which provides us with the best separation between MS-a
stars and the other MSs of $\omega$~Cen.  The reason could be that the
MS-a has a somehow peculiar CNO content with respect to the other
populations.  As above outlined, filters centered in the
blue-ultraviolet region, between $\sim$3200 and $\sim$4300 \AA, are
the most affected by CN and CH features (see Marino et
al. 2008, their Fig. 14).

Moreover, the He content must affect the position of the different
MSs, as discussed in previous sections.  Panel (a) of Fig.\ \ref{fig9}
shows the $m_{\rm F336W}$ vs.\ $m_{\rm F336W}-m_{\rm F435W}$ CMD of
$\omega$~Cen.  The MS-a fiducial sequence (drawn by eye) is plotted in
red in panel (b).  We subtracted the color of this fiducial sequence
from the color of all the stars at the same magnitude. The rectified
MSs are presented in panel (c) of Fig.\ \ref{fig9}.  We restricted our
analysis to the magnitude interval 19.7$<$$m_{\rm F336W}$$<$21.4,
where MS-a can be easily separated from the other MSs.  We drew a
vertical line, located at $\Delta(m_{\rm F336W}-m_{\rm F435W})=-0.065$
to isolate MS-a members (on the right) from the rest of the MSs [on
  the left, hereafter called (b+r)MS for simplicity].  We defined
three radial intervals in such a way that each radial bin contains the
same amount of selected stars [panels (d)].  For each of the three
radial intervals, using a logarithmic scale to emphasize MS-a counts,
panel (e) plots the distribution of the rectified colors. The ratios
of the star counts of MS-a/(b+r)MS are plotted in panel (f) as a
function of the angular distance from the cluster center.  Errors are
calculated as in Bellini et al.\ (2009): for each radial
interval, we derived the MS-a/(b+r)MS ratio in 5 magnitude bins [as
  defined by the red horizontal lines in panel (c)], and we used the
corresponding number of stars as weight to cumpute a weighted mean for
the MS-a/(b+r)MS ratio in each of the 5 bins.  Finally, we derived an
error for the entire radial interval from the residuals of the
individual ratio values from their mean, using the same weights as we
had used for the mean.  The radial trend shown in panel (f) of
Fig.\ \ref{fig9} is consistent, within the errors, with the flat
radial distribution of RGB-a stars (the progeny of the MSa stars) with
respect to (RGB-MInt+RGB-MP) ones (as found by Bellini et
al.\ 2009) within the inner $\sim$2 arcmin from the
cluster center.

\subsection{The sub-giant and lower red giant branches}

This region of the CMD was previously analyzed by Sollima et
al.\ (2005) and by V07.  The latter studied the $m_{\rm
  F435W}$ vs.\ $m_{\rm F435W}-m_{\rm F625W}$ ACS/WFC CMD.  They
identified four distinct stellar groups (named, from bright to faint
magnitudes, A, B, C and D, see Fig.\ \ref{fig2}) corresponding to at
least four distinct stellar populations, plus a broad distribution of
stars, between groups C and D.

WFC3 photometry reveals a new, much-more-complex picture of the SGB
region.  In the $m_{\rm F275W}$ vs.\ $m_{\rm F275W}-m_{\rm F814W}$ CMD
of Fig.\ \ref{fig10} and \ref{fig11}, stars of the original B and C
components of V07 are widely spread in the F275W band, without any
apparent substructure, while the brightest SGB component (A) is split
into two branches Figure \ref{fig12} shows a more quantitative
analysis of the split of SGB-A. The dashed red line has been traced
(by hand) between the two branches of SGB-A. For each star in the red
box which includes the dashed line, we calculated the difference
between the star magnitude and the magnitude of the dashed line at the
same color of the star.  The bimodal distribution of these magnitude
differences shown by the histogram in the lower part of
Fig.\ \ref{fig12} confirms the presence of two distinct branches.
 
It is not clear how the two SGB-A sequences evolve into the RGB,
though in the middle panels of Fig.\ \ref{fig11} they seem to run
parallel up to the bright part of the RGB.  In particular, the origin
of the bluest RGB is not obvious: is it coming from SGB-B or from the
faintest SGB of SGB-A?  The bluest RGB could also be something similar
to the broadened RGB of M4 visible in Fig.~11 of Marino et
al. (2008), where the broadening has been related to a
spread in CNO affecting the $U$-band.  Only chemical abundance
measurements will allow us to answer this question.  Interestingly
enough, the separation of the different RGB sequences becomes more
visible in the CMD $m_{\rm F275W}$ vs. $m_{\rm
F275W}-m_{\rm F814W}$ (see Fig.\ \ref{fig11}).

There is another feature of the SGB which is visible for the first
time in the CMDs presented in this paper.  The SGB-D of V07 (which
corresponds to the SGB-a Ferraro et al.\ 2004) is also
broadened, as shown in Fig.\ \ref{fig10}, \ref{fig11} and, in more
details, in Fig.\ \ref{fig13}.  It is not clear whether this
broadening corresponds to two distinct populations.  A visual
inspection of all these figures suggests that the faintest part of
SGB-D could be associated to a poorly populated MS which runs on the
red side of the MS-a.  The sequence on the red side of the MS-a cannot
be a sequence of binaries, which would evolve into a brighter (not
fainter) SGB.

In summary, the new WFC3 photometry shows that the SGB of $\omega$~Cen
is even more complex than thought so far. There are at least
\textit{six} distinct sequences, plus the broad distribution of stars
between SGB-C and SGB-D already identified by V07.

\subsection{The horizontal branch}

Typically, the horizontal branch (HB) amplifies all the complexities
of a stellar population, and it cannot be different for
$\omega$~Cen. Indeed, the HB shown in Fig.~\ref{fig14} shows a
multiplicity of features and, in particular, a well-known,
very-extended HB, with a pronounced blue hook (D'Cruz et
al.\ 2000).  In this section we want to focus the attention
on the blue hook.

The blue hook has a complex morphology, and it has been already
studied by Cassisi et al.\ (2009) and D'Antona, Caloi, \&
Ventura (2010), using the present ACS/WFC data set from
O-9442.  The new interesting feature displayed by the WFC3/UVIS data
set, and clearly shown in the inset of Fig.~\ref{fig14}, is that the
blue hook is split into two distinct, well-defined, separated -- and
almost vertical -- sequences.  The bluer blue-hook sequence contains
80$\pm$5\% of the total blue hook population, while the remaining
20$\pm$5\% of blue hook stars populate a redder parallel sequence
shifted by about 0.3 magnitudes in the $m_{\rm F275W}-m_{\rm F814W}$
color.

We also note that, on the red side of the two blue hooks (see
Fig.~\ref{fig14}), the HB seems to be separated into a fainter (more
populated) and a brighter component, up to at least $m_{\rm
  F275W}-m_{\rm F814W}=0.4$.

\section{Electronic catalog}
\label{sec_catalog}

The astro-photometric catalog will be available at the SIMBAD on-line
database\footnote{\texttt{http://simbad.u-strasbg.fr/simbad/}}.
Table~\ref{tab:cat} shows the first entries of the catalog.
Description of the catalog: column (1) contains stars ID; columns (2)
and (3) give the J2000.0 equatorial coordinates in decimal degrees;
columns (4) and (5) provide the pixel coordinates $x$ and $y$ of the
distortion-corrected reference meta-chip.  Columns from (6) through
(13) contain photometric measurements.  Note that the public catalog
gives the original photometry. The reddening and photometric zero
point spatial variation corrected photometry is available upon request
to the authors\footnote{Note that this work is based on images taken
  in July 2009. New WFC3/UVIS epochs have been (and others will soon
  be) collected for the same field of $\omega$~Cen analyzed here.}.

\begin{sidewaystable}
\centering 
\caption{First lines of the electronically available catalog.}
\label{tab:cat}
\scriptsize{
\begin{tabular}{ccccccccccccc}
\hline\hline
&&&&&&&&\\
ID$\!\!$&$\!\!$R.A. (J2000.0)$\!\!$&$\!\!$Dec. (J2000.0)$\!\!$&$\!\!$X$\!\!$&$\!\!$Y$\!\!$&$\!\!$$m_{\rm F225W}$ $\!\!$&$\!\!$ $m_{\rm F275W}$ $\!\!$&$\!\!$  $m_{\rm F336}$  $\!\!$&$\!\!$  $m_{\rm F435W}$$\!\!$&$\!\!$  $m_{\rm F606W}$$\!\!$&$\!\!$  $m_{\rm F625W}$ $\!\!$&$\!\!$   $m_{\rm F658W}$ $\!\!$&$\!\!$  $m_{\rm F814W}$\\
&&&&&&&&\\
\hline
&&&&&&&&\\
 1$\!\!$&$\!\!$201.7162676 $\!\!$&$\!\!$  $-47.5126489$$\!\!$&$\!\!$ $1781.197$$\!\!$&$\!\!$$426.673$$\!\!$&$\!\!$  23.421$\!\!$&$\!\!$   22.291$\!\!$&$\!\!$   21.088$\!\!$&$\!\!$   20.971$\!\!$&$\!\!$   19.990$\!\!$&$\!\!$   19.760$\!\!$&$\!\!$   19.526$\!\!$&$\!\!$   19.311\\
 2$\!\!$&$\!\!$201.7156870 $\!\!$&$\!\!$  $-47.5123958$$\!\!$&$\!\!$ $1809.429$$\!\!$&$\!\!$$444.904$$\!\!$&$\!\!$  20.674$\!\!$&$\!\!$   19.843$\!\!$&$\!\!$   19.095$\!\!$&$\!\!$   19.166$\!\!$&$\!\!$   18.436$\!\!$&$\!\!$   18.248$\!\!$&$\!\!$   18.056$\!\!$&$\!\!$   17.886\\
 3$\!\!$&$\!\!$201.7142821 $\!\!$&$\!\!$  $-47.5123325$$\!\!$&$\!\!$ $1877.748$$\!\!$&$\!\!$$449.480$$\!\!$&$\!\!$  21.045$\!\!$&$\!\!$   20.157$\!\!$&$\!\!$   19.368$\!\!$&$\!\!$   19.426$\!\!$&$\!\!$   18.650$\!\!$&$\!\!$   18.468$\!\!$&$\!\!$   18.271$\!\!$&$\!\!$   18.092\\
 4$\!\!$&$\!\!$201.7150112 $\!\!$&$\!\!$  $-47.5122876$$\!\!$&$\!\!$ $1842.291$$\!\!$&$\!\!$$452.709$$\!\!$&$\!\!$  24.545$\!\!$&$\!\!$   22.899$\!\!$&$\!\!$   21.604$\!\!$&$\!\!$   21.380$\!\!$&$\!\!$   20.289$\!\!$&$\!\!$   20.133$\!\!$&$\!\!$   19.908$\!\!$&$\!\!$   19.579\\
 5$\!\!$&$\!\!$201.7138235 $\!\!$&$\!\!$  $-47.5122802$$\!\!$&$\!\!$ $1900.054$$\!\!$&$\!\!$$453.253$$\!\!$&$\!\!$  23.926$\!\!$&$\!\!$   22.485$\!\!$&$\!\!$   21.130$\!\!$&$\!\!$   21.040$\!\!$&$\!\!$   20.033$\!\!$&$\!\!$   19.801$\!\!$&$\!\!$   19.578$\!\!$&$\!\!$   19.333\\
 6$\!\!$&$\!\!$201.7158384 $\!\!$&$\!\!$  $-47.5122430$$\!\!$&$\!\!$ $1802.055$$\!\!$&$\!\!$$455.906$$\!\!$&$\!\!$  21.261$\!\!$&$\!\!$   20.377$\!\!$&$\!\!$   19.555$\!\!$&$\!\!$   19.596$\!\!$&$\!\!$   18.831$\!\!$&$\!\!$   18.625$\!\!$&$\!\!$   18.421$\!\!$&$\!\!$   18.248\\
 7$\!\!$&$\!\!$201.7153042 $\!\!$&$\!\!$  $-47.5121646$$\!\!$&$\!\!$ $1828.040$$\!\!$&$\!\!$$461.554$$\!\!$&$\!\!$  21.184$\!\!$&$\!\!$   20.318$\!\!$&$\!\!$   19.513$\!\!$&$\!\!$   19.597$\!\!$&$\!\!$   18.803$\!\!$&$\!\!$   18.633$\!\!$&$\!\!$   18.426$\!\!$&$\!\!$   18.202\\
 8$\!\!$&$\!\!$201.7150457 $\!\!$&$\!\!$  $-47.5121237$$\!\!$&$\!\!$ $1840.615$$\!\!$&$\!\!$$464.507$$\!\!$&$\!\!$  21.843$\!\!$&$\!\!$   20.906$\!\!$&$\!\!$   20.033$\!\!$&$\!\!$   20.095$\!\!$&$\!\!$   19.282$\!\!$&$\!\!$   19.092$\!\!$&$\!\!$   18.866$\!\!$&$\!\!$   18.681\\
 9$\!\!$&$\!\!$201.7166455 $\!\!$&$\!\!$  $-47.5121043$$\!\!$&$\!\!$ $1762.811$$\!\!$&$\!\!$$465.879$$\!\!$&$\!\!$  21.047$\!\!$&$\!\!$   20.206$\!\!$&$\!\!$   19.413$\!\!$&$\!\!$   19.500$\!\!$&$\!\!$   18.726$\!\!$&$\!\!$   18.548$\!\!$&$\!\!$   18.340$\!\!$&$\!\!$   18.132\\
10$\!\!$&$\!\!$201.7164538 $\!\!$&$\!\!$  $-47.5120773$$\!\!$&$\!\!$ $1772.129$$\!\!$&$\!\!$$467.827$$\!\!$&$\!\!$  21.519$\!\!$&$\!\!$   20.593$\!\!$&$\!\!$   19.773$\!\!$&$\!\!$   19.820$\!\!$&$\!\!$   19.058$\!\!$&$\!\!$   18.870$\!\!$&$\!\!$   18.676$\!\!$&$\!\!$   18.444\\
\dots&\dots&\dots&\dots&\dots&\dots&\dots&\dots&\dots&\dots&\dots&\dots&\dots\\
&&&&&&&&\\
\hline
\hline
\end{tabular}}
\end{sidewaystable}

\acknowledgments
We thank S. Cassisi, F. D'Antona, and R. Gratton for useful discussions.
A.B.\ acknowledges support by the CA.RI.PA.RO.\ foundation, and by the
STScI under the {\it ``2008 graduate research assistantship} program.
G.P., A.F.M., and A.P.M.\ acknowledge partial support by MIUR under
the program PRIN2007 (prot.\ 20075TP5K9) and by ASI under the program
ASI-INAF I/016/07/0.

{}

\end{document}